\documentclass[12pt]{iopart}
\usepackage{graphicx,latexsym,stmaryrd,iopams}
 
\begin{document}

\title{Evidence for the super Tonks-Girardeau gas}

\author{M.T. Batchelor\footnote[3]{Corresponding author (Murray.Batchelor@anu.edu.au)}, 
M. Bortz, X.W. Guan and N. Oelkers
}
\address{
{\small Department of Theoretical Physics, RSPhysSE, and}\\
{\small Department of Mathematics, MSI}\\
{\small The Australian National University, Canberra ACT 0200, Australia}}


\begin{abstract}
We provide evidence in support of a recent proposal by Astrakharchik \etal for the existence 
of a super Tonks-Girardeau gas-like state in the attractive interaction regime of
quasi-one-dimensional Bose gases.  
We show that the super Tonks-Giradeau gas-like state corresponds to a highly-excited Bethe 
state in the integrable interacting Bose gas  for which the bosons acquire hard-core behaviour. 
The gas-like state properties vary smoothly throughout a wide range from strong repulsion to strong 
attraction. There is an additional stable gas-like phase in this regime in which the bosons form 
two-body bound states behaving like hard-core bosons. 
\end{abstract}

\maketitle

A sequence of striking advances has seen the experimental realization of quantum gases
of ultracold atoms in optical lattices.
Experiments in trapping one-dimensional ($1$D) Bose gases have been remarkably 
successful \cite{Exp-B1,Exp-B2,Exp-B3,Exp-B4,Exp-B5}. 
The effective  $1$D coupling constant $g_{1 \rm D}$ can be tuned essentially  from the weak  
coupling regime to the strong coupling regime via anisotropic confinement or Feshbach resonance.
The ratio of the average interaction energy to the kinetic energy, $\gamma=\frac{mg_{\rm 1D}}{\hbar^2n}$, 
is used to characterize the different physical regimes of the $1$D quantum gas.
Here $m$ is the atomic mass and $n$ is the boson number density.
For the regime in which the interactions are weakly repulsive, i.e., $\gamma\ll 1$, the $1$D
degenerate Bose gas acts like a quasi Bose-Einstein condensate.
As the strength of the repulsive interaction tends to infinity the bosons behave like
impenetrable fermions and the system is known as the Tonks-Girardeau gas. 
In this regime, the ground state properties of the $1$D homogeneous
Bose gas can be investigated using the wave function of
noninteracting fermions \cite{Girardeau} as a special limit of the Bethe Ansatz
solvable Bose gas  \cite{LL}.
On the other hand, observation of the formation of bright solitons in
quasi-$1$D trapped condensates \cite{Att-boson1,Att-boson2} has
renewed interest in the exactly solved $1$D model of interacting bosons 
in the {\em attractive} regime \cite{Att-boson3,Att-boson4}.
It was shown that attractive bosons confined to a finite system with
periodic boundary conditions can undergo a transition from uniform
condensation to bright solitons \cite{Att-boson4,extra,Att-boson7}.
These studies reveal subtleties in the signature of bosons in $1$D
many-body systems \cite{Bose1,Bose2,Bose3,Bose4,Bose5,Bose6,Bose7}.

It is very interesting to consider what happens to the $1$D Bose gas when the
coupling strength is abruptly changed from strong repulsive interaction to strong 
attractive interaction -- does the gas suddenly collapse or
does it inherit a stable gas-like phase from the hard-core bosons?
A recent proposal \cite{Astrak, report} by the group at {\rm BEC-INFM} in
Italy suggests the existence of a super Tonks-Girardeau gas for 
the strongly attractive regime.
They argued that this new gas-like state can be stable in a wide range of strongly 
attractive interaction strength. 
In using a variational Monte Carlo calculations they found that this gas-like
state can be described by a gas of hard rods with strong attraction,
which yields a positive compressibility.
In this regime the $1$D Bose gas is more strongly correlated than the
Tonks-Girardau gas -- they thus referred to it as the {\em super Tonks-Girardau gas}.
Here we show that this super Tonks-Girardau gas-like state corresponds to a 
highly-excited state in the Bethe Ansatz solution of the interacting Bose gas for 
strongly attractive interaction.
We also predict the existence of another stable gas-like phase where the
bosons form two-body bound states with large binding energy.
The existence of these gas-like states is due to the fact that a large
Fermi-pressure-like kinetic energy inherited from the Tonks-Girardeau gas
cannot be quenched instantly when the interaction is switched from
strong repulsion to strong attraction.
This is analogous to a system in classical mechanics where a moving
ball is able to reside on a step of a declining staircase in a potential well if the 
friction between the ball and the surface of the steps is strong enough.
In addition, for weakly attractive interaction the bosons form a
many-body bound state which leads to the ground state energy per
particle being proportional to the linear density $n$.

The interacting $\delta$-function Bose gas \cite{LL,MC} is described by the Hamiltonian
\begin{equation}
{H}=-\frac{\hbar ^2}{2m}\sum_{i = 1}^{N}\frac{\partial ^2}{\partial x_i^2}+\,g_{\rm 1D} 
\sum_{1\leq i<j\leq N} \delta (x_i-x_j)
\label{Ham-1}
\end{equation}
in which $N$ particles are constrained by periodic boundary conditions
on a line of length $L$. 
Here $g_{\rm 1D} ={\hbar ^2 c}/{m}$ is an effective 1D coupling constant with scattering strength $c$. 
If the gas is harmonically trapped along the axial direction, the scattering strength is  
determined by $c={2}/{|a_{\rm 1D}|}$, 
where $a_{\rm 1D}$ is the effective 1D scattering length.
The eigenvalues are given by ${E}=\frac{\hbar^2}{2m}\sum_{j=1}^Nk_{j}^2$, where the
quasi-momenta $k_j$ satisfy the Bethe equations
\begin{equation}
\exp(\mathrm{i}k_jL)=- \prod^N_{\ell = 1} 
\frac{k_j-k_\ell+\mathrm{i}\, c}{k_j-k_\ell-\mathrm{i}\, c}, \qquad \mbox{for} \quad j = 1,\ldots, N.
\label{BE}
\end{equation}

The Bethe roots $k_j$ are known to be real for repulsive interactions,
$c > 0$, where the model has been extensively studied 
(see also, Refs.~\cite{KOR,Tbook,Wadati,BGM,BGNL,LUTT1} and references therein).
A significant hallmark of $1$D bosons is that the
quasi-momentum distribution satisfies a universal semi-circle law for
relatively weak interaction strength \cite{LL,gau71,BGM}. 
As the interaction strength increases the distribution function becomes flat and the bosons
behave like impenetrable free fermions.  
More generally, the long-wavelength properties of the 1D Bose gas can be
described by a Luttinger liquid due to the universality in the low-energy excitations, i.e., 
gapless excitations with a linear spectrum in the low-energy excitations and power-law decay 
in the correlations \cite{Haldane,CAZA1,LUTT1}.

On the other hand, the situation is far from completely understood for attractive interactions, 
$c<0$, where the roots form bound states in the complex plain.
It is believed that for attractive interaction there is no saturation
due to the absence of a hard core in the strong coupling limit.
McGuire \cite{MC} argued that the ground state energy for
$N$ particles in infinite space is $E_0=-\frac{1}{12}c^2N(N^2-1)$, 
which agrees with the result later obtained from the string solution \cite{Tbook}. 
This implies that there is no thermodynamic limit. 
However, Astrakharchik \etal \cite{Astrak, report} have suggested the existence of a
new stable gas-like phase in the 1D Bose gas for strongly attractive interaction. 
Using variational Monte Carlo calculations they found that for a certain range of strong coupling, 
the energy per particle coincides with the energy 
\begin{equation}
\frac{E}{N}=\frac{\pi^2\hbar^2n^2}{6m}\left(1+\frac{2n}{c}\right)^{-2} \label{hardrods}
\end{equation}
of a gas of hard rods.
This leads to a positive compressibility. 
In this regime there exists an energy barrier which separates this gas-like state from cluster-like
bound states \cite{Astrak2}. 
It was also shown that the short range correlations for the super Tonks-Girardeau gas-like phase 
are much stronger than in the usual Tonks-Girardeau gas.

Before discussing the strongly attractive regime, consider first the weakly attractive case.
Here we see that the exactly solved 1D model of attractive interacting bosons 
does saturate and the ground state energy for the gas phase changes smoothly as the
interaction varies from weakly repulsive to weakly attractive. 
Following \cite{BGM}, we find that the roots of the Bethe
equations for  weakly attractive bosons satisfy  a set of algebraic BCS-like equations of the form
\begin{equation}
k_j=\frac{2\pi d_j}{L}+\frac{2c}{L} \, {\sum_{\ell=1}^{N}}'\frac{1}{k_j-k_\ell}, \quad j=1,\ldots, N 
\label{momenta}
\end{equation}
where the summation excludes $j=\ell$ and $d_j=0,\pm 1,\pm 2, \ldots$ denote excited states.
The ground state has zero total momentum, with $d_j=0$ for $j=1,\ldots,N$, with the roots
forming an $N$-body bound state.
These equations are the same as for the weakly repulsive regime
\cite{gau71,BGM}, although now the roots of equation (\ref{momenta})
are all purely imaginary for the ground state as $c < 0$.
They are given by Hermite polynomials of degree $N$, i.e., $H_N(\sqrt{\frac{L}{2c}}k)=0$ \cite{gau71}. 
Alternatively, the roots may be given in terms of Laguerre polynomials \cite{BGM}.
The ground state energy per particle is ${E_0}/{N}={c(N-1)}/{L}$, 
which follows directly from (\ref{momenta}) (see also Ref.~\cite{SSAC}). 
In this way the ground state energy per particle in the weakly attractive and repulsive regimes is 
seen to pass continuously through the origin at $c=0$. 
Since our analysis is based on finite systems, it avoids the subtlety of a singular integration 
kernel in the Lieb-Liniger integral equation in the limit $c\to 0$.
Our findings agree with the previously reported numerical results 
for the attractive regime \cite{SSAC,Att-boson5,Att-boson6}.

An alternative approximation of the Bethe equations for weak 
interaction consists in observing that in this regime, each 
configuration of the Bethe-numbers can be described by $N_c$ clouds of 
quasi-momenta, centred around $k_j^{(0)}=2\pi j/L$, where integer $j\neq 0$ corresponds to excited states. 
The momenta of each cloud lie parallel to 
the imaginary (on the real) axis for $c<0$ ($c>0$). More quantitatively, 
let us label the $k$'s by two indices $j$ and $\nu$, where 
$j=1,\ldots,N_c$ and $\nu$ counts the particles $n_j$ in each cloud $j$, 
$\nu=1,\ldots,n_j$. Then
\begin{eqnarray}
k_{j,\nu}&=& k_j^{(0)} +\delta_j +\beta_\nu^{(j)}, \label{k}
\end{eqnarray}
where $k_j^{(0)}$ denote the corresponding quasimomenta for $c=0$. 
Here $k_j^{(0)} +\delta_j$ can be viewed as the ``centre of mass" of the 
$j$-th cloud in quasi-momentum space and $\beta_\nu^{(j)}$ describes the 
position of the individual quasimomenta within one cloud around its centre of mass.
By linearizing the Bethe equations including $O(c)$ terms, one obtains
\begin{eqnarray}
\delta_j&=&\frac{2 c}{L} \sum_{\ell\neq j} \frac{n_k}{k_j^{(0)}-k_\ell^{(0)}}\label{delta}\\
\beta_\nu^{(j)}&=& \frac{2 c}{L} \sum_{\mu\neq \nu} 
\frac{1}{\beta_\nu^{(j)}-\beta_\mu^{(j)}} \label{beta}
\end{eqnarray}
Note that according to (\ref{beta}), the momenta within one cloud are 
given by roots of Hermite polynomials, centred around the cloud centre of mass. 
Only this centre of mass is influenced by the presence of other clouds, cf. (\ref{delta}).  
Thus the linearization of  the Bethe equations is analogous to a  dipole expansion in 
electrostatics, with charges  placed at the position of the quasimomenta. 
Note that Eqs.~\eref{delta} and \eref{beta} coincide with Eq.~\eref{momenta} only for the ground state. 
The reason is that for excited states,  Eq.~\eref{momenta} contains terms of order higher than $O(c)$.
It turns out that Eq.~\eref{momenta} yields a better approximation for $c<0$, 
whereas Eqs.~\eref{delta} and \eref{beta} are better suited for $c>0$.
It is straightforward to calculate the energy $E_{cloud}$ from Eqs.~\eref{delta} and \eref{beta},
\begin{eqnarray}
E_{cloud}=\sum_{j=1}^{N_c} \left[ p_j^{(0)}\right]^2 n_j + \frac{c}{L} N(2N-1) 
- \sum_{j=1}^{N_c} n_j^2. \label{energy}
\end{eqnarray}

As an example, consider $N=12$ weakly interacting bosons.
Tables 1 and 2 show a direct comparison between the ground state energy and the 
leading eight excitations obtained from the various equations. 
These are the energies obtained directly from numerical solution of the Bethe equations \eref{BE}, 
the BCS-like equations \eref{momenta} and the cloud equations \eref{k}-\eref{energy}.
Also shown are the energies obtained from the direct
diagonalization  \cite{Att-boson4} of the truncated Hamiltonian.
The overall agreement is excellent, with even very close energy values
clearly distinguishable by their characteristic root distributions and total momentum.

The momentum density distribution for the ground state is along the imaginary axis, 
i.e., the Bethe roots are of the form $\mathrm{i}k_j$ for $j=1,\ldots, N$, 
and given by the semi-circle law 
$ n(k)=\frac{1}{\pi\sqrt{\gamma}}\sqrt{1-\frac{k^2}{4|\gamma| n^2}}$.
This is analogous to the behaviour in the weakly repulsive case \cite{LL,gau71,BGM}.
The momentum distribution also becomes flat if the interaction strength is increased.
The stronger the interaction strength, the larger the momentum distribution region.
This implies a signature for the 1D Bose gas in the weak attractive regime $\gamma \ll 1/N$.

We turn now to strongly attractive interaction. 
From the exact analytic result for two bosons with attractive delta interaction, 
if $c<\!-2$, the roots for the ground state tend to $\mathrm{i}\frac{c}{2}$. 
Without loss of generality, we focus on $N=2M$ bosons with $Lc\ll -1$ for even $M$. 
Then for general $N$, 
the roots of the Bethe equations \eref{BE} for the ground state are approximately 
given by \cite{MC,Tbook} 
$k_{\pm j}\approx \pm \mathrm{i}\left(\frac{c}{2}(2M-2j+1)+\delta_j\right),\,j=1,\ldots,M,$ 
where $\delta _j$ is small and negligible for $Lc\ll -1$.  
In this case the wave function is approximately given by \cite{Att-boson7}
\begin{equation}
\psi(x_1,\ldots,x_N)\approx{\cal N}\exp\left\{\frac{c}{2}\sum_{1\leq i<j\leq N}|x_j-x_i|\right\}.
\end{equation}
Here ${\cal N}=\frac{\sqrt{(n-1)!}}{\sqrt{2\pi}}|c|^{\frac{(n-1)}{2}}$ is a normalization constant. 
The eigenvalue of the Hamiltonian for this state is $E_0=-\frac{1}{12} c^2 N(N^2-1)$, which is 
McGuire's soliton-like ground state energy \cite{MC}.

On the other hand, if the interaction strength is abruptly changed from strongly repulsive
to strongly attractive, the gas-like phase may vary smoothly against the 
cluster-like phase \cite{Astrak,Astrak2}. 
We see this in the context of the exactly solved model from a highly-excited state in which
all Bethe roots are real and symmetric about the origin, with
$\pm k_{2m-1} ,\,m=1,\ldots, {N}/{2}$, where $k_{\ell} \approx \frac{\pi \ell}{L} 
\left(1+\frac{2}{\gamma} \right)^{-1} $.
These roots can be directly obtained from the asymptotic expansion of the Bethe 
equations \eref{BE} for $L|c| \gg 1$.
The energy of this state follows directly from this asymptotic solution, with
\begin{equation}
\frac{ E}{N} \approx \frac{\hbar^2 }{2m}\frac{1}{3}(N^2-1)
\frac{\pi ^2}{L^2}\left(1+\frac{2n}{c}\right)^{-2} \label{E0}
\end{equation}
which coincides with the result \eref{hardrods} for the gas of hard-rods \cite{Astrak,Astrak2}.
Here the condition $\frac{2N}{L|c|}\ll 1$ is required. 
For harmonic trapping along the axial and the radial directions, this condition becomes
$N\ll L/a_{\rm 1D}$, where $L$ is proportional to the characteristic
length $a_z =\sqrt{\hbar/m\omega_z}$, i.e. $L\propto \sqrt{N}a_z$. 
Therefore the strong coupling condition is equivalent to a restriction,  $N\ll a_z^2/a_{\rm 1D}^2$, 
on the number of bosons, which is similar to the requirement for the formation of bright solitons 
in attractive condensates \cite{Att-boson1,Att-boson2}. 
Here $\omega_z$ is the trapping frequency along the axial direction.
We obtain the sound velocity $v_s=v_{F}/(1+\frac{2}{\gamma})^2$ 
which is positive for strongly attractive coupling. 
This indicates that the compressibility \cite{Astrak} of the system is positive,
corresponding to a stable gas-like phase.
The Fermi velocity is $v_F=\frac{\hbar \pi n}{m}$. 
The Luttinger liquid parameter characterizing the density correlation
rate is $K=(1+\frac{2}{\gamma})^2$, indicating an enhancement of
the density-density correlations in this super Tonks-Girardeau gas.
For example, the one-particle correlations 
$g(x) = \langle \psi^\dagger(x) \psi(0) \rangle \propto 1/x^{\frac{1}{2K}}$.
In order to see the enhancement of the correlation function in the super 
Tonks-Girardeau phase, we write $K = (1-|a_{\rm 1D}| n)^2$, which is
smaller than in the Tonks-Girardeau phase, where $K = (1+|a_{\rm 1D}| n)^2$.

The local two-body correlation function
$g^{(2)}=4\pi^2n^2/3\gamma^2(1+\frac{2}{\gamma})^3$ measures the
probability of collisions between two particles.  
The interaction energy of the ultracold bosons depends on coherent collisions. 
In the strong coupling limit, the probability of collisions is very small due to 
decoherence of the wave functions, i.e., $g^{(2)}\to 0$. 
Therefore the interaction weakly affects the energy per particle in the
strong coupling regime due to the existence of kinetic pressure. 
Figure 1 shows a plot of the energy per particle \eref{E0} and the velocity versus the gas
parameter $n a_{\rm 1D}$.
We see that they vary smoothly as the interaction varies from strongly repulsive to
strongly attractive.
The existence of Fermi-pressure-like kinetic energy 
is the main reason for a stable gas-like state throughout the whole of the crossover
from strong repulsion to strong attraction.

Nevertheless, there  is  another stable gas-like state where the Bethe roots form 
two-body bound states, namely with complex conjugate pairs of the form
$\pm k_{2m-1}\approx \alpha_m \pm \mathrm{i}\left(\frac{c}{2} +\delta_{m}\right) ,\,m=1,\ldots, \frac{M}{2}$, 
with
\begin{eqnarray}
\alpha_{m} &=&\frac{(2m-1)\pi}{2L} \left(1+\frac{5M}{2Lc} \right)^{-1}\\[.25cm]
\delta_{m}&=&\frac{-2^{M-2}c^Me^{\frac{c}{2}L}}{\alpha_m^M\prod_{l=1}^{M/2}|(\alpha_m^M)^2-(\alpha_l^M)^2|+2^{M-2}Lc^Me^{\frac{c}{2}L}}.\label{del}
\end{eqnarray}
The case $l=m$ is excluded in the product in (\ref{del}).  
The energy per particle follows as
\begin{eqnarray}
\frac{ E}{N} \approx \frac{\hbar^2 }{2m}\left\{ -\frac{c^2}{4}+\frac{\pi^2(M^2-1)}{12L^2(1+\frac{5M}{2Lc})^{2}}-\frac{2^{M}c^{M+1}e^{\frac{c}{2}L}(1+\frac{5M}{2Lc})^{M-1}}{\frac{N\pi^{M-1}}{(2L)^{M-1}}\Gamma(M)}\right\}.
\end{eqnarray}
Here $\Gamma$ denotes the gamma function. 
In the RHS of the above equation the first term is the pair binding energy. 
The second term contains the kinetic energy and interaction energy. 
The last term is the correction from $\delta_m$ in \eref{del}. 
This correction term is negligible in the thermodynamic limit.  
Thus we have
\begin{eqnarray}
\frac{E}{N} \approx \frac{\hbar^2 n^2 }{2m}\left\{-\frac{\gamma^2}{4}+\frac{\pi^2}{48}
\left(1+\frac{5}{4\gamma}\right)^{-2}\right\}
\end{eqnarray}
This state is an analogue of the fully-bound fermion pairs in the
$1$D Fermi gas where the bound pairs cannot condense due to Fermi
pressure \cite{BEC-BCS1,BEC-BCS2,BEC-BCS3,BEC-BCS4,BBGN}. 
Those pairs can be thought of as hard-core bosons. 
For this state the sound velocity is effectively given by 
$v_s=\frac14 v_{F}(1+\frac{5}{4\gamma})^{-2}>0$, implying the stability of this
gas-like state. 
The Luttinger liquid parameter is $K=4(1+\frac{5}{4\gamma})^2$. 
The enhancement of $K$ in this phase leads to a very slow decay of the phase 
correlations \cite{Haldane,CAZA1}, but less strongly correlated than the super
Tonks-Girardeau gas.  
The sound velocity decreases with the formation of multi-body bound states.
The vanishing of the sound velocity implies that the system
is unstable in multi-soliton-like states where the system lacks
hard-core behaviour. 
But some soliton-like states which lead to positive compressibility would be stable. 
This is a manifestation of the formation of matter-wave soliton
trains in quasi-$1$D attractive condensates \cite{Att-boson1,Att-boson2}.

We conclude that some gas-like states could be stable against
cluster-like states if the interaction strength abruptly changes 
from strong repulsion to strong attraction. 
The gas can inherit hard-core behaviour due to the existence of 
large Fermi-pressure-like kinetic energy from the strongly repulsive interaction. 
This regime could potentially be 
reached in experiments via anisotropic confinement in which a
resonance of the effective $1$D scattering length takes place.

\ack
This work has been supported by the Australian Research Council and
the German Science Foundation under grant number BO-2538/1-1.

\clearpage

\begin{table}
\caption{
The ground state energy per particle ($j=0$) and the first nine excitations ($j=1,\ldots,9$) 
for $N=12$ weakly interacting bosons with $c=-0.05$ and $L=6$. 
Also shown is the total momentum $P$ of each state.
${E}_\textrm{\tiny exact}$ denotes the energies obtained directly from numerical solution of the 
Bethe equations \eref{BE}.
${E_\textrm{\tiny BCS}}$ denotes energies obtained from solutions of
the BCS-like equations \eref{momenta}.
${E_\textrm{\tiny cloud}}$ denotes energies obtained from solutions of
the cloud equations \eref{k}-\eref{energy}.
For further comparison,  
${E_\textrm{\tiny num.}}$ denotes energies obtained from direct
diagonalization of the truncated Hamiltonian.
}
\vskip 3mm
\begin{center}
\begin{tabular}{|c||c|c|c||c|c|}
\hline 
$j$&
$P$&
${E}_\textrm{\tiny exact}$&
${E_\textrm{\tiny BCS}}$& 
${E_\textrm{\tiny cloud}}$&
${E_\textrm{\tiny num.}}$\\
\hline 
$0$&
$0$&
$-0.11028$&
$-0.11000$&
$-0.10000$&
$-0.11021$
\\
\hline
$1$&
$ {\pi}/{3}$ &
$0.96786$&
$0.96809$&
$0.96829$ &
$0.96794$
\\
\hline
$2$&
$0$&
$2.04772$&
$2.04791$&
$2.04825$ &
$2.04782$
\\
\hline
$3$&
$ {2\pi}/{3}$&
$2.04936$&
$2.04955$&
$2.04991$&
$2.04946$
\\
\hline
$4$&
$ {\pi}/{3} $&
$3.13094$&
$3.13109$&
$3.13153$&
$3.13433$
\\
\hline
$5$&
$ \pi $&
$3.13422$&
$3.13438$&
$3.13487$&
$3.13432$
\\
\hline
$6$&
$ 0$& 
$4.21587$&
$4.21599$&
$4.21649$&
$4.21599$
\\
\hline
$7$&
$ {2\pi}/{3} $&
$4.21751$ &
$4.21764$ &
$4.21816$&
$4.21763$
\\
\hline
$8$&
$ {4\pi}/{3} $&
$4.22244$&
$4.22260$&
$4.22316$&
$4.22256$
\\
\hline
$9$&
${2\pi}/{3}$&
$4.25788$&
$4.25810$&
$4.25816$&
$4.25797$
\\
\hline
\end{tabular}
\end{center}
\end{table}

\begin{table}
\caption{
Same as for Table 1, but with $L=24$.}
\vskip 3mm
\begin{center}
\begin{tabular}{|c||c|c|c|c|c|}
\hline 
$j$    &
$P$  &
${E}_\textrm{\tiny exact}$&
${E_\textrm{\tiny BCS}}$&
${E_\textrm{\tiny cloud}}$&
${E_\textrm{\tiny num.}}$
\\
\hline
0 & 0 & $-0.02779$ & $-0.02750$ & $-0.02750$ &  $-0.02772$ 
\\
\hline
1 & ${\pi}/{12}$ & $0.03601$ & $0.03624$ & $0.03646$ & $0.03609$ 
\\
\hline
2 & $0$ & $0.10029$ 
& $0.10047$ & $0.10083$ & $0.10039$ 
\\
\hline
3 & ${\pi}/{6}$ & $0.10067$ & $0.10086$ & $0.10125$ &   $0.10077$
\\
\hline
4 & ${\pi}/{12}$ & $0.16542$ & $0.16557$ & $0.16603$ &   $0.16554$ 
\\
\hline
5 & ${\pi}/{4}$ & $0.16619$ & $0.16635$ & $0.16687$ &   $0.16631$
\\
\hline
6 & $0$ 
&  $0.23102$& $0.23114$ & $0.23166$ &   $0.23115$
\\
\hline
7 & ${\pi}/{6}$ & $0.23141$ & $0.23153$ & $0.23207$ &    $0.23154$
\\
\hline
8 & ${\pi}/{3}$ & $0.23258$ & $0.23271$ &  $$0.23332&    $0.23270$
\\
\hline
9 & ${\pi}/{6}$ & $0.24178$ & $0.24202$ & $0.24207$ &    $0.24188$
\\
\hline
\end{tabular}
\end{center}
\end{table}

\clearpage

\begin{figure}[t]
\begin{center}
\vskip 5mm
\includegraphics[width=.70\linewidth]{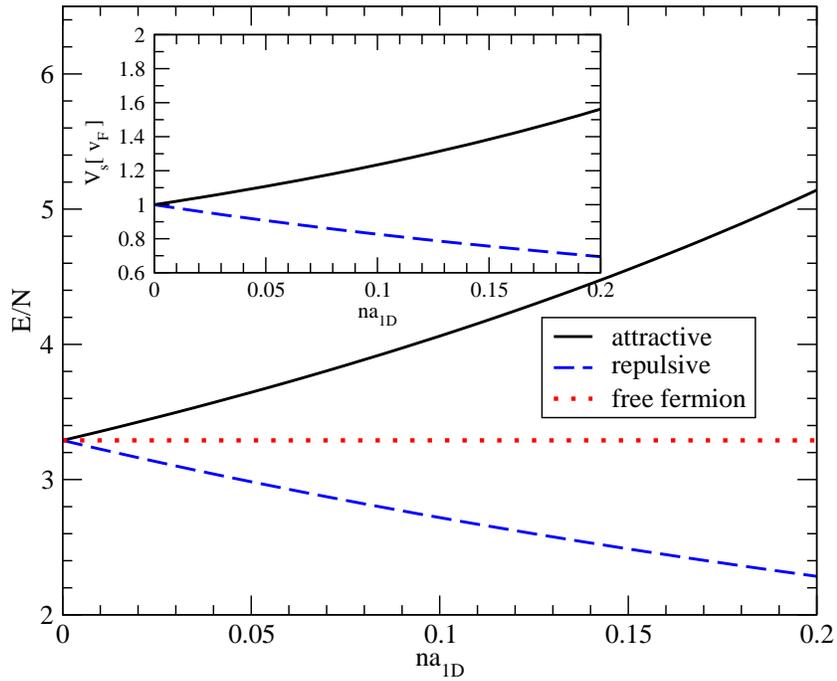}
\end{center}
\caption{The energy per particle in units of $\hbar^2 n^2 /2m$ versus the gas parameter $na_{\rm 1D}$
for strong attractive and strong repulsive interaction. Here $g_{\rm
1D}=2\hbar^2/ma_{\rm 1D}$.  The inset shows the sound velocity versus
the gas parameter. The gas-like state varies smoothly due to the
existence of Fermi-pressure-like kinetic energy if the interaction strength 
is abruptly changed from strongly repulsive to strongly attractive.}
\label{fig:SE}
\end{figure}

\end{document}